\documentstyle[12pt,epsfig,rotating]{article}
\def\bd{
\begin{document}} \def\ed{\end{document}}
\def\emp{\end{minipage}} \def\bmp{\begin{minipage}}
\def\bcc{\begin{center}} \def\ecc{\end{center}}     \def\npg{\newpage}
\def\beq{\begin{equation}} \def\eeq{\end{equation}} \def\hph{\hphantom}
\def\be{\begin{equation}} \def\ee{\end{equation}} \def\r#1{$^{[#1]}$}
\def\n{\noindent} \def\ni{\noindent} \def\pa{\parindent} 
\def\hs{\hskip} \def\vs{\vskip} \def\hf{\hfill} \def\ej{\vfill\eject} 
\def\cl{\centerline} \def\ob{\obeylines}  \def\ls{\leftskip}
\def\underbar#1{$\setbox0=\hbox{#1} \dp0=1.5pt \mathsurround=0pt
   \underline{\box0}$}   \def\ub{\underbar}    \def\ul{\underline} 
\def\f{\left} \def\g{\right} \def\e{{\rm e}} \def\o{\over} \def\d{{\rm d}} 
\def\vf{\varphi} \def\pl{\partial} \def\cov{{\rm cov}} \def\ch{{\rm ch}}
\def\la{\langle} \def\ra{\rangle} \def\EE{e$^+$e$^-$}
\def\bitz{\begin{itemize}} \def\eitz{\end{itemize}}
\def\btbl{\begin{tabular}} \def\etbl{\end{tabular}}
\def\btbb{\begin{tabbing}} \def\etbb{\end{tabbing}}
\def\beqar{\begin{eqnarray}} \def\eeqar{\end{eqnarray}}
\def\\{\hfill\break} \def\dit{\item{-}} \def\i{\item} 
\def\bbb{\begin{thebibliography}{9} \itemsep=-1mm}
\def\ebb{\end{thebibliography}} \def\bb{\bibitem}
\def\bpic{\begin{picture}(260,240)} \def\epic{\end{picture}}
\def\akgt{\noindent{Acknowledgements}}
\def\fgn{\noindent{\bf\large\bf Figure captions}}
\bd

\vskip-6.5cm
\hskip12cm{\large HZPP-9904}

\hskip12cm{\large July 25, 1999}

\vskip0.5cm

\centerline{\Large  Model Investigation of Non-Thermal Phase Transition}
\medskip
\centerline{\Large  in High Energy Collisions}
\medskip
\medskip

\centerline{ Wang Qin \hskip1cm Li Zhiming \hskip1cm Liu Lianshou }
\medskip

\centerline{\small (Institute of Particle Physics, Huazhong Normal University,
Wuhan, 430079) }
\vskip3cm

\centerline{\large Abstract  }
\medskip
\begin{center}\begin{minipage}{104mm}
{\small \hskip1cm  The Non-thermal phase transition in high energy collisions
is studied in some detail in the framework of random cascade model.
The relation between the characteristic parameter $\lambda_q$ of phase
transition and the rank $q$ of moment is obtained
using Monte Carlo simulation, and the existence of 
two phases in self-similarly cascading multiparticle systems is shown.
The relation between the critical point $q_c$ of phase transition 
on the fluctuation parameter $\alpha$ is obtained and compared with
the experimental results from NA22. The same 
study is carried out also by analytical calculation under central
limit approximation. The range of validity of the central limit
approximation is discussed.\\}
\end{minipage}
\end{center}
\vskip2cm

{\large  Keywords\ \ \  random cascade \  multifractal \  
           anomalous scaling \  

\hskip3cm non-thermal phase transition   }

\newpage
Recently, the prediction\r{1} that there exist the property of self-affine 
fractal in the anisotropic phase
space of multiparticle final states in high energe hadron-hadron 
collisions has been confirmed by experiments\r{2,3}.
This breakthrough  in the nonlinear study of high energy physics 
places the further study of nonlinear property
of multiparticle final states on the agenda.

In this respect, the non-thermal phase transition\r{4,5} is a prolem 
worthy while further study. In the presently available
experiments\r{6}, due to the restriction of energy,
the average multiplicity is very low, and the rank 
of the factorial moments could not be high. So, 
no clear evidence of non-thermal phase transition has been seen. The
new Large Hadron Collider (LHC), which is being built and will be put 
into operation in the beginning of next century, will dramatically
raise the collision energy and multiplicity, providing perfect condition 
for the study of non-thermal phase transition.  For a theoretical 
preparation it is necessary to carry on detailed discussion on this
phase transition and to clarify its property.

The aim of this short paper is to make a model study of the non-thermal 
phase transition, especially to make clear of the relation between the 
critical point of non-thermal phase transition and the strength of 
dynamical fluctuations.

The random cascading $\alpha$ model is  widely used in the study of nonlinear 
property of multiparticle final states in high energy collisions. 
Using this model, it is easy to  
get a system pocessing the  property of intermittency and fractal.
We will show that non-thermal phase transition does exist in this
system and the relation between the critical 
point of phase transition and the parameter $\alpha$ of 
fluctuation strength in the model can thus be obtained and
compared with the experimental data.

Firstly, let us briefly  remind 
the random cascading $\alpha$ model\r{7} with probability conservation.

Consider a region $\Delta$ of one-dimensional phase space.
Devide  it into  $\lambda$ cells.
The probability of particles falling into the $i$th cell is
   $$p_i=p_0\omega_i , \eqno(1)$$
where $p_0=1$ is the probability in the phase space region $\Delta$,
$\omega_i$ is the probability of the elementary partition. 
Next, we divide each sub-bin into $\lambda$ even smaller sub-bins.
The probability in the $i j$th bin  $(i=1,2, \dots, \lambda ; \ j=1,2,\dots,
\lambda)$  is
  $$p_{ij}=p_i \omega_j  ,   \eqno (2) $$
After  $\nu$ steps, the probability in a sub-bin is 
$$p_{i_1 i_2  \cdots  i_ \nu}= \prod_{k=1}^ \nu \omega_{i_k} .\eqno(3) $$
The total number of intervals is $ M= \lambda^ \nu $.      

In order to guarantee the conservation of probability  in each step of 
cascading, we choose the elementary probability $\omega_i$ 
for $\lambda=2$ as\r{7}:
 $$\omega_1=\frac{1+\alpha r}{2}  ,\qquad
\omega_2=\frac{1-\alpha r}{2}. \eqno(4)$$ 
where $r_i$ is a random number distributed uniformly in the 
interval $[-1,1]$,
$\alpha$ is a model parameter describing the strength of nonlinear 
dynamical fluctuations ($0<\alpha<1$). 

The definitions of the probability moments and factorial moments are\r{8}:
 $$  C_q= \frac{ \frac{1}{M} \la \sum_{m=1}^M p_m^q \ra }
{ \f( \frac{1}{M} \la \sum_{m=1}^M p_m \ra \g)^q  }
= M^{q-1} \la \sum_{m=1}^M p_m ^q\ra .  \eqno(5) $$
  $$ F_q\f(M\g)= \frac{1}{M} \sum_{m=1}^M \frac{ \la n_m \f(n_m-1 \g) 
  \cdots \f(n_m-q+1 \g)\ra}{{\la n_m \ra}^q} . \eqno(6) $$
It can easily be proved that under the assumption of Poisson or Bernoulli 
type of statistical fluctuations the normalized factorial moments $F_q$ are 
equal to the normalized probability moments $C_q$.

The character of dynamical fluctuations can be expressed as the anomalous 
scaling of probability (or factorial) moments:
   $$ C_q \f(M \g) \propto M^ {\varphi_q} ,  $$   
or equivalently 
$$ \ln C_q(M )= A+\varphi_q \ln M  \qquad
 \f(M\rightarrow \infty \g) .\eqno(7) $$
where $\varphi_q$ is called intermittency index. 

In order to see the anomalous scaling of probability moments more clearly,
we choose the fluctuation-strength parameter $\alpha=0.5$, 
the elementary partition number $\lambda=2$, the division step $\nu =12$,  
the ranks of moment  $q=5,10,15,20,25,30 $,  
and make use of Eq.(5) to simulate the relation $\ln C_q{\sim} \ln M $.
The results are shown in Fig.1.  
The  intermittency parameters $ \varphi_q$ are obtained through
linear fit.  We can see from the figure that 
the higher the rank $q$ is, the larger the slope  
$\varphi_q$ is.

A  parameter $\lambda_q$ has been introduced \r{4,5} in the multifractal 
analysis to characterise the non-thermal phase transition in the 
multiparticle systems.  It is related to the intermittency index 
$ \varphi_q$ by the relation
    $$ \lambda_q= \f( \varphi_q+1 \g)/q . \eqno(8) $$
We will try to evaluate this parameter both through analytic calculation and 
by using Monte Carlo simulation.

In the random cascading $\alpha$ model, the probability moment is: 
 $$C_q \f(M \g)= \frac{\la \omega^q \f(1 \g) \cdots \omega ^q \f(\nu \g) \ra}
  {{\la \omega \ra}^{q \nu} } . \eqno(9) $$
It can be rewritten as:
 $$C_q \f(M \g)=\lambda^{q \nu} \la \omega^q \f(1 \g) \cdots \omega ^q 
  \f(\nu \g) \ra $$ 
 $$ =\lambda^{q\nu}\f\la \exp\f(-\sum_{i=1}^\nu q\varepsilon_i \g)\g\ra , 
\eqno(10)$$
where $\varepsilon_i=-\ln \omega \f(i \g)$. The parameter  
$\zeta=\sum_{i=1}^\nu q\varepsilon_i$ in the above equation is the sum of 
$\nu$ random numbers. Under the central limit approximation
 $\zeta$ approades to Gaussian distribution:
  $$ C_q \f(M \g)= \lambda^{q \nu} \la \e^{-q \zeta} \ra =\exp \f(q \nu 
\ln \lambda +\frac{\nu \sigma^2q^2}{2}-q \bar{\zeta} \g) . \eqno(11) $$
 Using $C_1(M)=1$, we get
   $$ C_q \f(M \g)=\e^{\nu \sigma^2q \f(q-1 \g)/2} . \eqno(12) $$
The intermittency indices can be deduced as:   
   $$ \varphi_q= \frac{\f(q-1 \g)q \sigma^2}{2 \ln2} .  \eqno(13) $$ 
We have also the relation  
$$ \sigma^2=\la {\ln}^2\omega \ra-{\la \ln \omega \ra}^2=
\frac{1}{3}\sigma^2+\frac{2}{3}\sigma^4+\cdots\cdots  \ . $$
Under linear approximation\r{9} it becomes $\sigma^2=\alpha^2 / 3 $.
Substituting into Eq.(13) we get
   $$ \varphi_q=\frac{q \f(q-1 \g) \alpha^2}{6\ln2} .  \eqno(14) $$    
from eq.(8):
  $$ \lambda_q=\frac{\varphi_q+1}{q}=\frac{\f(q-1 \g)\alpha^2}
 {6\ln2}+\frac{1}{q} \ \ .  \eqno(15)$$
The resulting  $\lambda_q {\sim}q$ are plotted in Fig.2($a$)
for $\alpha =0.2 ,0.3 ,0.4 ,0.5$ respectively.

Fig.2($a$) is the result under central limit approximation.
The exact relation can not be calculated analytically. 
Therfore, we use Monte Carlo simulation.  The resulting 
$\lambda_q {\sim}q$ for $\alpha =0.2 ,0.3 ,0.4 ,0.5$ respectively, 
are shown in Fig.2$(b)$.

It can be seen from the figures that the $\lambda_q {\sim}q$ curves 
from both the Monte Carlo simulation and the analytical calculation under
central limit approximation have the same trend, i.e. with the 
increasing of $q$, $\lambda_q$ arrive at a minimum at the point $q_c$, 
which means that there really exists non-thermal phase transition in the 
self-similar cascading model and two different phases do indeed coexist,
$q_c$ is the critical point of phase transition.

In Fig.2 we alse draw the experimental data from NA22 
(open circles). It stops at the rank $q=5$ and is unclear
whether there is a minimum at some higher rank as required by
non-thermal phase transition. 
The open triangles in the figure is the result from the same experiment
selecting only the particles with low transverse momenta
($p_t < 0.15$ GeV/$c$). In this case, with the increasing of $q$
(from 4 to 5), $\lambda_q$ increases. It seems to show that there is phase 
transition and the critical point $q_c < 5$. 
As is well known, choosing only the particles with low transverse 
momenta, the strength of intermittency increases\r{10}. 
Therefore, this experimental phenomenon shows that the system  
with lower transverse momenta, which has larger intermittency strength, 
has lower critical point of non-thermal phase transition. 
This is qualitatively the same as 
the result of our model, where the phase transition point shifts left 
with the increasing of fluctuation strength
(when $\alpha$ increases $q_c$ decreases). 

In order to see more clearly the relation between 
the phase transition point $q_c$  
and the fluctuation parameter $\alpha$ of the model, we draw the figure of  
$q_c  {\sim} \alpha$, as shown in Fig.3.  We can see from the figure that
the larger $\alpha$ is,  the earlier $q_c$ appears.

Comparing the exact values of $q_c  {\sim} \alpha$
from Monte Carlo simulation
and the analytical results under central limit approximation,
it can be seen that both have the same trend of continuously descending.
However, the values of $q_c$ in central limit approximation are
generally smaller than the exact values.
This shows that the central limit approximation can reflect
qualitatively the property of non-thermal phase transition
but there is noticeable quantitative deviation.

\newpage

\begin{picture}(260,240)
\put(-110,-460)
{\epsfig{file=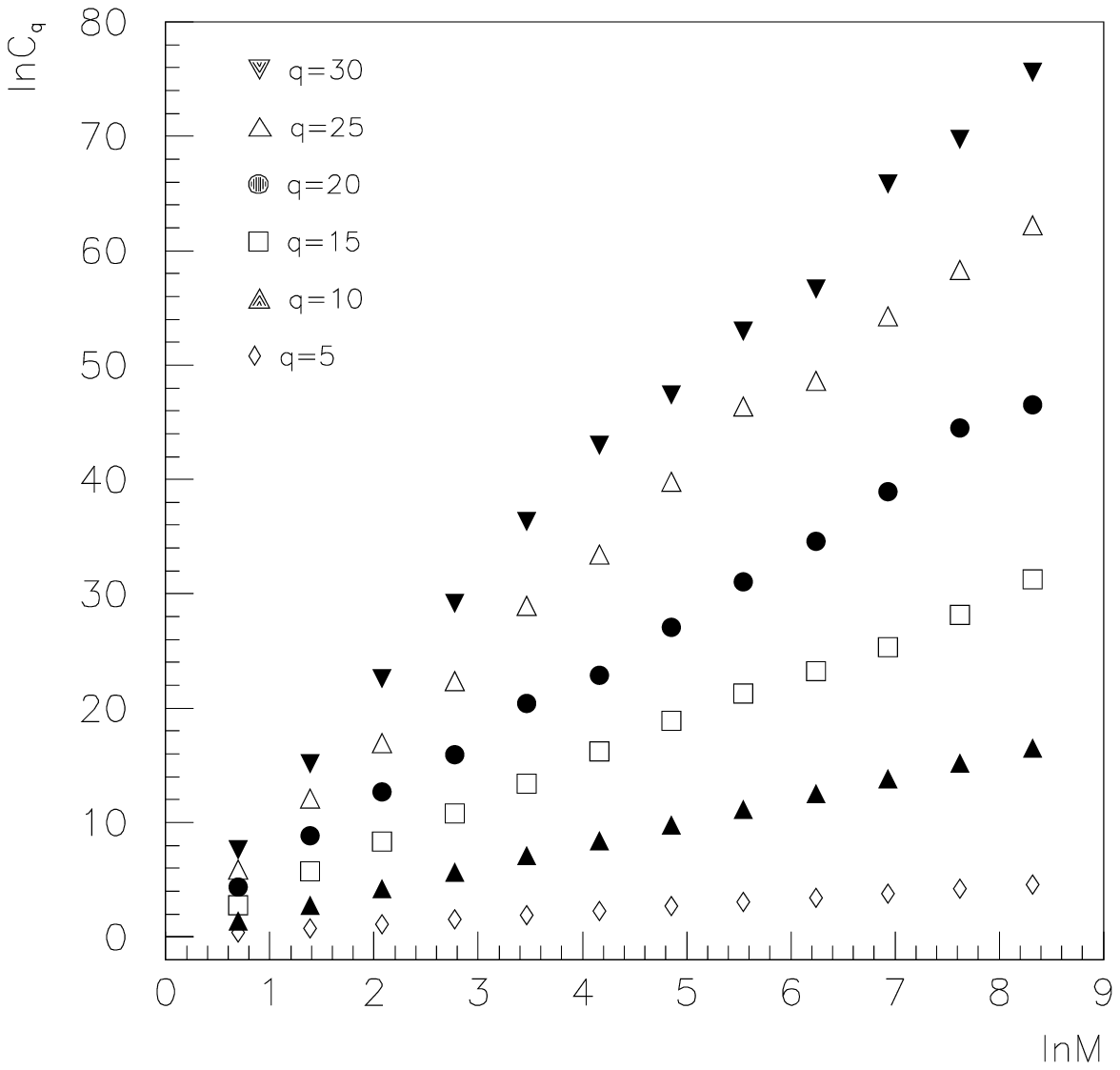,bbllx=0cm,bblly=0cm,
           bburx=8cm,bbury=6cm}}
\end{picture}

\vs8cm
\n{Fig.1 \ \ Log-log plot of various rank probability moments versus 
partition number in $\alpha$ model}

\newpage

\begin{picture}(260,240)
\put(10,-410)
{\epsfig{file=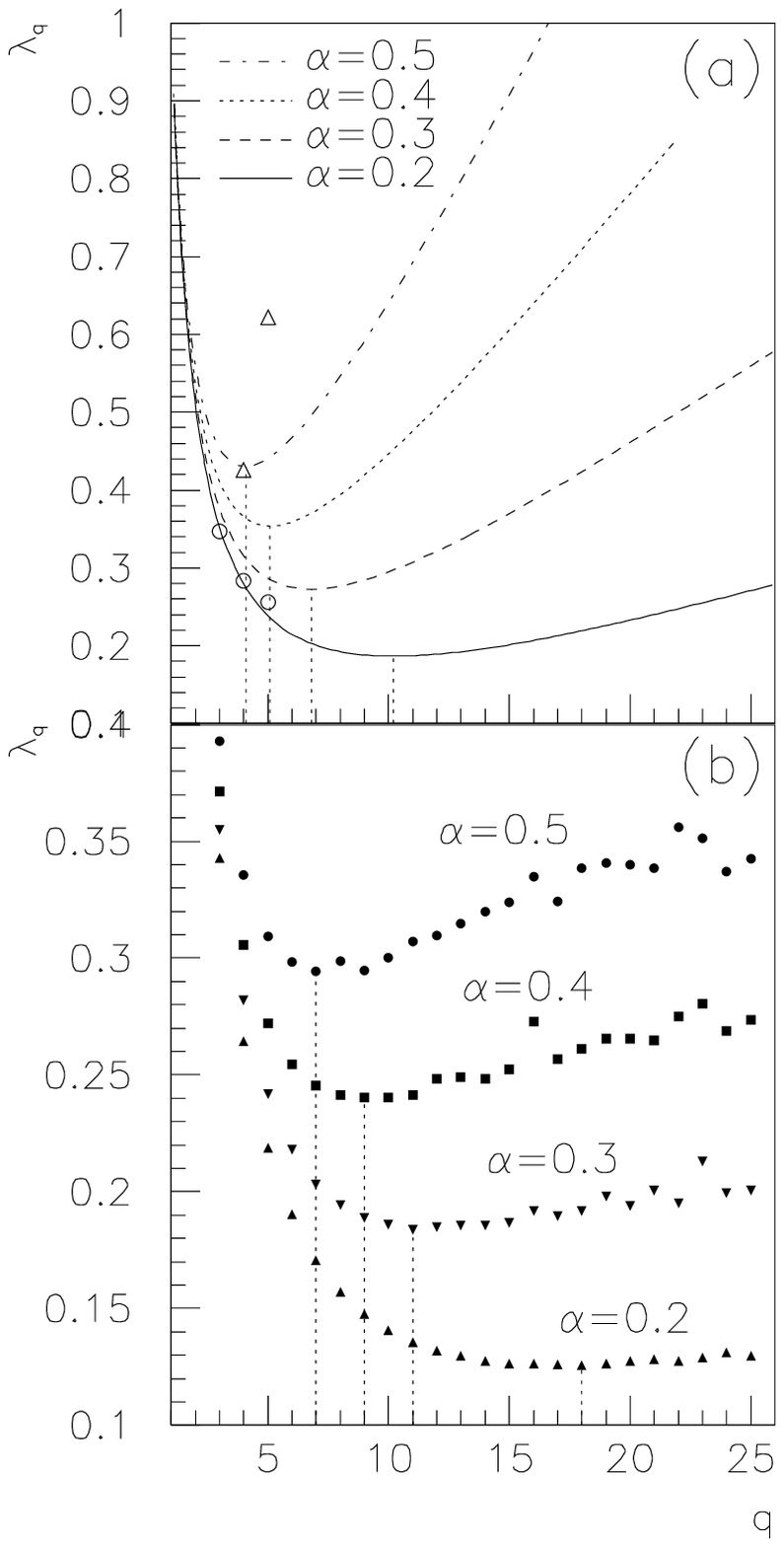,bbllx=0cm,bblly=0cm,
           bburx=8cm,bbury=6cm}}
\end{picture}

\vs9cm
\n{Fig. 2 \ \ 
Relation between the parameter $\lambda_q$ and the rank $q$ of moments.
The vertical lines indicate the position of minima.} \ \ 
($a$) \ Analytical results under central limit approximation.
The open circles are the experimental results from NA22.
Open triangles are the results from the same experiments 
taking only low momentum particles with 
($p_t < 0.15$ GeV/$c$) Data taken from Ref.[6]. \ \ 
($b$) \ Results of Monte Carlo simulation.
\newpage

\begin{picture}(260,240)
\put(-110,-460)
{\epsfig{file=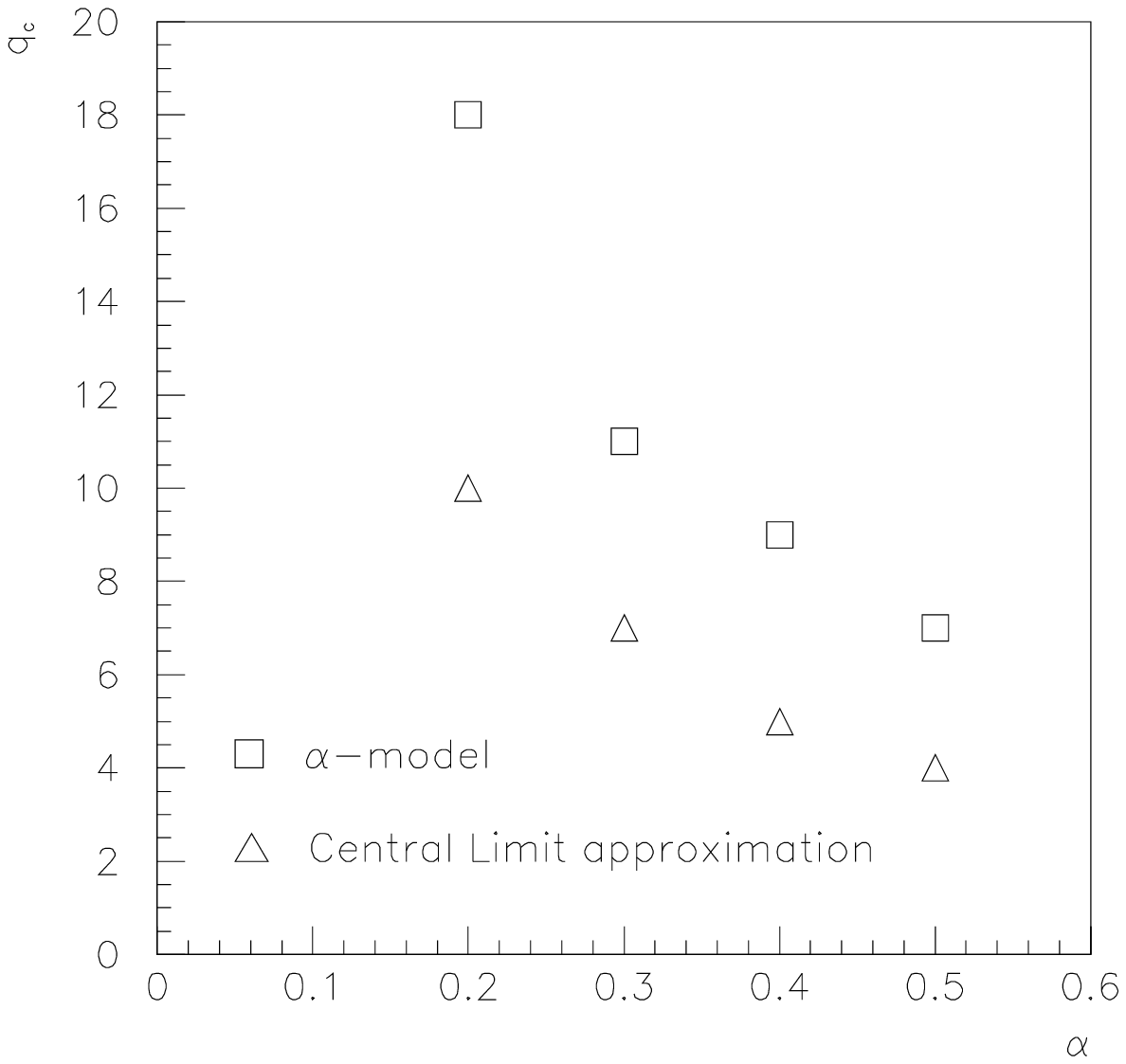,bbllx=0cm,bblly=0cm,
           bburx=8cm,bbury=6cm}}
\end{picture}

\vs8cm
\n{Fig. 3 \ \ The relation between phase trnasition point $q_c$ 
and fluctuation strength  $\alpha$}

\ed